# Human-Data Interaction: Thinking beyond individual datasets

**Authors:**
Laura Koesten, University of Vienna
Jude Yew, Google
Kathleen Gregory, Centre for Science and Technology Studies, Leiden University

Highlights:
- Shifting perspective from data use to data interaction
- Documenting dynamic contexts: Towards data as processes
- Co-locating data and use facilitates data interaction

**Abstract:**
Having greater access to data leads to many benefits, from advancing science to promoting accountability in government to boosting innovation. However, merely providing data access does not make data easy to use; even when data is openly available online, people may struggle to work with it. In this article, we draw on prior work, including our own, and a case study of Kaggle – a large online data science community – to discuss the importance of moving away from viewing datasets as static resources. Instead, we describe the view of data as a *process* with its own interactional affordances that offer many different possibilities for data, as well as for social interaction. We advocate for the notion of *Human-Data Interactions* and their potential implications for various audiences.

**Introduction**

We live in data-centric times. Many of the world's greatest challenges, from advancing science and improving government services to tackling climate change, all require access to large amounts of data. In recent years, there has been a proliferation of avenues for individuals to share and use data online. Scientists use open platforms such as GitHub or Zenodo to collaborate on data projects and archive them. Governments establish data hubs for publishing and facilitating data use, or policymakers and industry develop "data spaces" for safe data sharing across multiple parties. There are also more informal platforms to share data (and code) within the data science community, such as Kaggle[1] for machine learning competitions or Hugging Face[2] with a focus on natural language processing technologies.

However, using data created by others is still challenging. Just providing access to data does not ensure that it is discoverable or usable; attention needs to be paid to how data is interacted with as well. Despite the proliferation of data-sharing websites and platforms, there is still a lack of understanding about the social interaction, collaboration, and questioning found around data. We argue that attention should be paid to these and other aspects of ``human-data interaction'' – focusing on the issues and affordances of how people actually use and interact with data.

---

[1] https://www.kaggle.com/
[2] huggingface.co/datasets

In this article, we draw on insights learned from our analysis of data practices on Kaggle[3], which involved a questionnaire investigating data discovery and evaluation behaviours and observations of platform capabilities and metadata. We also draw on prior work by ourselves and others (e.g.;Faniel et al., 2019; Birnholtz et al.,2003; Koesten & Gregory et al., 2021) to illustrate how data interaction could be facilitated and to suggest the idea of data as a process.

We start with Kaggle as an example of diverse data interactions. Then, we structure our argument around four themes, addressing factors related to the data (such as descriptions and quality conceptions) and factors related to engagement (like discussion forums and co-located tools). We conclude by offering recommendations for data publishers, platforms, and the data science community, emphasizing the shift from seeing data as static objects to facilitate more fluid data interactions.

**Kaggle as a use case**
Founded in 2010, Kaggle has now over 16 million users. Initially focusing on machine learning competitions, it has transformed into a platform for data science learning and collaboration. Users exchange datasets and code, collaborate on projects, and compete in data science challenges.
As of January 2024, more than 288,000 public datasets of different formats and more than 5000 competitions have been hosted on Kaggle. The platform offers integrated notebooks for collaborative cloud-based data analysis, enabling users to create, share, and collaborate on code, data, and text. They can also browse, filter, and search datasets based on various criteria like file size, type, license, and tags. Kaggle is an interesting case study for several reasons:
- It is the largest platform of its kind.
- It is host to a substantial data science community of practice, beyond competitions.
- It directly supports data work on the platform itself
- Other platforms such as GitHub have a distinct focus i.e. code or data publishing.
- It implements a diverse range of user-centric features.

Kaggle offers a collaborative data work setting where users share data and tools, collaborate on analyses, and sometimes team up to compete. Unlike other online communities like peer production systems, citizen science or Q&A sites, there's limited literature on data platforms. We use Kaggle as an example to explore Human-Data Interactions, highlighting the use of data beyond its original purpose.

**From data use to data interaction**
Traditionally, data use has been seen to consist of first finding data, then downloading it and working on it locally, possibly in a team. In this context, a dataset is understood as a resource shared by someone online or in a repository with the purpose to make it accessible to others. As we see on Kaggle, the ways such reuse happens is shifting towards more fluid practices. Data, code, documentation, and reviews are available in the same place in the cloud; this allows potential users to try out data on the fly, talk to others about it, and make more confident decisions about its fitness of use for new tasks.

Kaggle offers an example of how such an environment could be designed. Our analysis illustrates how the design of data sharing/reuse platforms has an impact on data use itself. Of the most popular datasets on Kaggle, most have co-located code and long, diverse discussions attached to them, suggesting that combining data, code, documentation, and discussion can lead to a higher degree of data use. We also see that Kaggle, and similar platforms, cater not just to experts; rather, they open up data science practices to a wider range of learners, by providing more real-time context, descriptions and interactions with community members.

---

[3] https://zenodo.org/records/10998910

**Dynamic contexts**
These types of discussions and community interactions are one way of providing context to data. Research has repeatedly shown that datasets cannot be isolated from their context without losing meaning (Faniel et al., 2019). They are situated within a platform, a community, and norms of production and use (Neff et al., 2017). Context is also important in meeting what can be thought of as an individual's 'data needs' which are shaped by combinations of relationships between roles, professional norms, and intended uses for data (Gregory & Koesten, 2022).

Clearly defining context as it pertains to data remains challenging (Faniel et al., 2019), although there is general consensus that data use requires contextual documentation (Birnholtz and Bietz, 2003; Borgman, 2015). This includes information *from* the data (such as relevance, usability, and quality criteria), as well as information *about* the data, for example where and how it was created (Gregory & Koesten, 2022). Contextual details can be in the form of metadata, information about the reputation of the publishing institution, repository or individual, indications of prior data use, as well as social context provided via user interactions. We observed that context is also documented via interactions on Kaggle, where people review and discuss data just like any other digital resource. These interactions provide the social context of that data - its popularity, how frequently it is used, etc. Kaggle enables dataset ranking through community feedback and provides integrated notebooks for browser-based data exploration. These features make visible the dynamic nature of context.

**Documenting dynamic contexts: Towards data as processes**
On platforms such as Kaggle, datasets are used in varied ways, e.g. as they are cleaned, merged, and then put back online. Data is changing as it is updated and extended, but also as it is being used in other contexts. For instance, joining tables to produce new datasets that will find new uses perhaps. This suggests that describing data once at the point of creation or publication might not suffice. Metadata and context also shift when data takes on different meanings in different data use scenarios. This can become particularly challenging when integrating data from diverse sources and levels of granularity, as different contexts collide. Transformations of data often remain hidden. Some transformations, e.g. the remixing of different data or the details of what gets added or removed during different uses can potentially be traced automatically. However, changes in context and meaning often remain invisible and are difficult to document.

The different questions that data is used to answer might change the meaning of the data beyond that for which it was originally created. Literature examining the reuse of scientific data illustrates the extent to which data is influenced by contexts of creation and use, which define what counts as data, what gets disregarded, and how data is organised and processed to fit various purposes (Borgman, 2015). Using data for different purposes requires transformation and remixing - not just of data variables, but a remixing of context and meaning. Current metadata schemas arguably lack elements necessary to capture these more interactional elements. These different types of changes – of meaning, context, and to the structure of data – foreground the idea of what we refer to as "data as a process," demonstrating how data is continually changing as it is being used.

How these changes can be documented in a meaningful way is a key question for the future of data reuse. This includes and requires not just interactions with the data but also social interactions, such as asking questions, engaging in a dialogue around the data as well as social signals. e.g. in the form ratings. These social interactions and conversations themselves make the dynamic nature of data visible. However, in many established data repositories, e.g, in science or government, such conversations are less supported and also less common on the platforms themselves.

Some platforms in different environments have adapted to the need of collaboratively building up contextual knowledge and adapting content, such as for open source code. Kaggle offers some possibilities for how dynamic contexts of data can be documented, but we believe there are other interesting avenues to consider. Just like audio or video, data should be easily findable, flexibly partitioned, remixed, repurposed, and shared as part of the usage cycle, beyond notebooks and metadata, discussions, and documentation. We should be able to find, reuse, provide attribution, and track the process of evolution of a dataset as we can a Wikipedia article, for example.

Current data use practices, and tools linked to them, feel too static and rigid to create the same type of effects we have noticed in other user-generated content that have democratized how content is produced and consumed online. One new possibility for data could be to document different levels of reuse, depending on the number of times data is transformed based on prior analysis efforts. To some extent versioning also allows us to keep track of such interactions. Another possibility to make changes in data visible could be to indicate how a particular dataset was derived -- i.e. by showing and summarizing the differences between the reused dataset and its original source alongside documentation to facilitate more fluid usage of data, going beyond existing efforts for data summaries (e.g.; Phillips and Smit, 2021).

**Co-locating data and use**
We know that there are many different ways of working with data, due in part to the size of the data and storage requirements. However, we believe that bringing data and their uses together by making processes more transparent can support human data interactions.

This can involve working with data in the cloud - e.g. where data is dynamically updated in notebooks or code. It can also involve thinking about how to make data work more transparent for data that is not (or cannot be) openly shared. Shared data analyses of confidential data, at an aggregate level, are one way to make data use more visible. These analyses could be co-located with openly available metadata. Even if the data cannot be shared, data elements and traces of uses could be co-located. Bringing together data and its different uses could also help address challenges like inaccessible data due to missing or broken download links and the spread of data across multiple locations.

Kaggle is mainly used for learning exploratory data analysis and creating prototype solutions[4]. However, this model does not directly apply to complex production environments that require debugging, testing, and deployment workflows integrated into existing IT infrastructure. While notebook environments are useful for initial and small-scale projects, they may not fully handle the complexity of real-world data science projects. Overall, we see the role of Kaggle as not just a competition platform, but rather as a hub to gain access to useful learning resources and expertise in data science. Learning environments might not directly transfer to professional environments in which data are used in new contexts. At the same time, the use of Kaggle as a place to find and experiment with data, often in teams, resonates with other data-sharing platforms, such as GitHub, open government, and open science portals (Koesten et al., 2020; Sicilia et al., 2017).

**Implications**
Thinking about the importance of human data interaction and viewing data as a process has concrete implications for data publishers, as well as for environments supporting data science practices and data teamwork. We believe Kaggle is in many ways an interesting example beyond its specific purpose, as many data portals are re-considering their goals and efforts to shift from publishing more data to supporting use and building communities. Within organizations (or across them, along supply chains and partner networks), the use of data is playing an increasing role. The challenges are in many ways similar to those of public platforms - there are decisions to be made about what data to share with other business units or partners and how to design an environment where people can make sense of shared datasets, and put them to use. Viewing data as a process can help shape our thinking about how communities can form around datasets, how sensemaking as an iterative process can be supported by interaction design, and how data reuse itself is dynamic.

---
[4] https://zenodo.org/records/10998910

*Implications for data publishers*
Our arguments have direct implications for those developing data publishing software or managing existing data publishing programs. We advocate for a more user-centric idea of context, and particularly data quality to make it easier for people to judge whether a dataset fits their purpose. We recommend other parties involved in the supply of data to experiment with platform capabilities, including:
- Encourage interaction/community development by defining clear tasks or questions provided alongside a dataset, open to a community who can engage around these tasks.
- Allow and encourage community ratings, such as votes.
- Set up infrastructure and tools to capture and analyze user engagement and interaction flows.
- Consider meaningful data use indicators and metrics, assess data use continuously, and act on the insights.
- Provide rich documentation and other resources to describe data, including text explanations, key column headers and their expected value ranges, known issues, etc.

Data providers have arguably so far missed the opportunity to use approaches and techniques that have proven their value time and time again in other, genuinely consumer-oriented domains, from online search to retail to user-generated content platforms. Our understanding of what makes data easier to use is still limited, especially when it comes to tangible evidence on the role of various publishing guidelines and platform affordances on engagement (Koesten et al., 2020).

*Implications for supporting human data interaction in data science environments:*
Looking at data science practices, many professionals are not trained data scientists, but rather increasingly have to use data on the job and need to gain more specialised skills to be productive. In that regard, Kaggle as a community of data science learners is not that different from any organisation adapting their staff skills to meet demands raised by digitalization and remote work. We recommend organisations undergoing such transformations to:

- Facilitate learning through shared resources and community engagement.
- When designing for dataset discovery, such as in an enterprise data lake, remember that users have varied search criteria based on their data needs, especially inexperienced users.
- Co-locate data and tools, i.e. to enable ease of understanding, access, and real-time feedback.
- Provide descriptions and explanations of datasets (e.g., a dataset's biography, including accounts of prior usage and example code).
- Invest in cleaning key datasets to allow learners to focus on data exploration and analysis.

Open data platforms can play a role in training the next generation of data practitioners. In many ways, Kaggle looks like an online community of practice, but there isn't much research into how existing frameworks and insights on how learning and apprenticing in practice-based communities apply to data science. As much data science education happens online, there is a wealth of experience and best practices to draw from to inform the design of data publishing platforms.

*Implications for data science teamwork:*
This applies to the broader context of remote data-centric work, with dispersed team members and diverse expertise. The scale of problems and infrastructure requires us to think about solutions that support people with a range of backgrounds and expertise, working together with minimal direct contact. This and similar efforts prepare organizations for future challenges in remote data science work. Our recommendations for organizations using data science are:

- Facilitate discussions on access and visibility levels.
- Recognize diverse data needs within teams.
- Provide a platform for multiple perspectives to converge on understanding data risks, value, affordances, and ethical considerations.
- Consider competitions as a framework to encourage teamwork and learning.